\documentclass[11pt,draftcls,onecolumn]{IEEEtran} 
\usepackage{latexsym, amssymb, amsmath}
\usepackage{graphicx}
\usepackage{epsfig}

\usepackage{url}
\usepackage{algorithmic}
\usepackage{algorithm}
\usepackage{array}
\usepackage{eqparbox}
\usepackage[tight, footnotesize]{subfigure}

\usepackage{bm}

\newcommand{\calN}{\mathcal{N}}
\newcommand{\by}{\mathbf{y}}

\newcommand{\bP}{\mathbf{P}}
\newcommand{\be}{\mathbf{e}}
\newcommand{\bV}{\mathbf{V}}
\newcommand{\bU}{\mathbf{U}}
\newcommand{\ba}{\mathbf{a}}

\newcommand{\bc}{\mathbf{c}}

\newcommand{\SNR}{\mathrm{SNR}}
\newcommand{\calM}{\mathcal{M}}
\newcommand{\calE}{\mathcal{E}}

\newcommand{\bt}{\mathbf{t}}
\newcommand{\btheta}{\bm{\theta}}
\newcommand{\mathdef}{\stackrel{\triangle}{=}}
\newcommand{\transpose}{\mbox{${}^{\text{T}}$}} 
\newcommand{\Frac}[2]{{{#1}/{#2}}}  

\newenvironment{remark}[1][Remark]{\begin{trivlist}
\item[\hskip \labelsep {\bfseries #1}]}{\end{trivlist}}

\begin{document}


\title{Estimating Signals with Finite Rate of Innovation from Noisy Samples:\\ 
  A Stochastic Algorithm}

\author{Vincent Y.~F. Tan,~\IEEEmembership{Graduate Student Member,~IEEE,} Vivek~K~Goyal,~\IEEEmembership{Senior Member,~IEEE}  
\thanks{Vincent Y.~F. Tan (vtan@mit.edu) is with the Massachusetts Institute of
Technology, 77 Massachusetts Avenue, Rm. 32-D570, Cambridge, MA 02139,
Tel: 617-253-3816, Fax: 617-258-8364. Vincent Tan is supported by the Agency for Science, Technology and Research (A*STAR), Singapore. }
\thanks{Vivek K Goyal (vgoyal@mit.edu) is with the Massachusetts Institute of
Technology, 77 Massachusetts Avenue, Rm. 36-690, Cambridge, MA 02139,
Tel: 617-324-0367, Fax: 617-324-4290.}}

\maketitle
\begin{abstract}
As an example of the recently-introduced concept of rate of innovation,
signals that are linear combinations of a finite number of Diracs per unit time
can be acquired by linear filtering followed by uniform sampling.
However, in reality, samples are rarely noiseless.
In this paper, we introduce a novel \emph{stochastic} algorithm to reconstruct a signal with finite rate of innovation from its \emph{noisy} samples.
Even though variants of this problem has been approached previously, satisfactory solutions are only available for certain classes of sampling kernels, for example kernels which satisfy the Strang--Fix condition.
In this paper, we consider the infinite-support Gaussian kernel, which does not satisfy the Strang--Fix condition.
Other classes of kernels can be employed.
Our algorithm is based on Gibbs sampling, a Markov chain Monte Carlo (MCMC) method.
Extensive numerical simulations demonstrate the accuracy and robustness of our algorithm. 
\end{abstract}


\begin{center}
\bfseries EDICS Categories: DSP-SAMP, SSP-PARE
\end{center}

\IEEEpeerreviewmaketitle
\newpage
\section{Introduction}
\label{sec:Intro}

The celebrated Nyquist-Shannon sampling theorem~\cite{Sha, Nyq}%
\footnote{A more expansive term could be the Whittaker-Nyquist-Kotelnikov-Shannon sampling theorem; see, e.g., \cite{Jer, Uns}.}
states that a signal $x(t)$ known to be bandlimited to $\Omega_{\max}$ is uniquely determined by samples of $x(t)$ spaced $1/(2\Omega_{\max})$ apart.
The textbook reconstruction procedure is to feed the samples as impulses to
an ideal lowpass (sinc) filter. 
Furthermore, if $x(t)$ is not bandlimited or the samples are noisy,
introducing pre-filtering by the appropriate sinc \emph{sampling kernel} gives
a procedure that finds the orthogonal projection to the space of
$\Omega_{\max}$-bandlimited signals.
Thus the noisy case is handled by simple, linear, time-invariant processing.

Sampling has come a long way since the sampling theorem, but until recently the results have mostly applied only to signals contained in shift-invariant subspaces~\cite{Uns}. Moving out of this restrictive setting, Vetterli {\it et al.}~\cite{Vet02} showed that it is possible to develop sampling schemes for certain classes of non-bandlimited signals that are not subspaces. As described in~\cite{Vet02}, for reconstruction from samples it is necessary for the class of signals to have \emph{finite rate of innovation} (FRI).
The paradigmatic example is the class of signals expressed as
$$
  x(t)=\sum_{k} c_k \phi(t-t_k)
$$
where $\phi(t)$ is some known function.
For each term in the sum, the signal has two real parameters $c_k$ and $t_k$.
If the density of $t_k$s (the number that appear per unit of time) is finite,
the signal has FRI\@.
It is shown constructively in~\cite{Vet02} that 
the signal can be recovered from (noise-less) uniform samples of $x(t) * h(t)$
(at a sufficient rate) when $\phi(t) * h(t)$ is a sinc or Gaussian function.
Results in~\cite{Dra07} are based on similar reconstruction algorithms and greatly reduce the restrictions on the sampling kernel $h(t)$.

\begin{figure}[!t]
\centering
\includegraphics[height=1.2in]{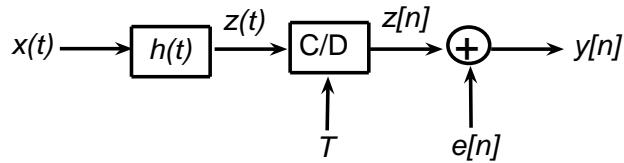}
\caption{Block diagram showing our problem setup. $x(t)$ is a signal with FRI given by (\ref{eqn:fri}) and $h(t)$ is the Gaussian filter with width $\sigma_h$ given by (\ref{eqn:kernel}). $e[n]$ is i.i.d.\ Gaussian noise with standard deviation $\sigma_e$ and $y[n]$ are the noisy samples. From $y[n]$ we will estimate the parameters that describe $x(t)$, namely $\{c_k,t_k\}_{k=1}^K$, and $\sigma_e$, the standard deviation of the noise.}
\label{fig:block}
\end{figure}

In practice, though, acquisition of samples is not a noiseless process. For instance, an analog-to-digital converter (ADC) has several sources of noise, including thermal noise, aperture uncertainty, comparator ambiguity, and quantization~\cite{Wal}. Hence, samples are inherently noisy. This motivates our central question: \emph{Given the signal model ({\it i.e.} a signal with FRI) and the noise model, how well can we approximate the parameters that describe the signal?} In this work, we address this question and develop a  novel algorithm to reconstruct the signal from the noisy samples, which we will denote $y[n]$ (see Fig.~\ref{fig:block}). 

\subsection{Related Work and Motivation}
Signals with FRI were initially introduced by Vetterli {\it et al.}~\cite{Vet02}. The reconstruction schemes hinged on identifying algebraically-independent parameters of the signals, {\it e.g.} the weights $\{c_k\}$ and time locations $\{t_k\}$. In the seminal paper on FRI, the sampling kernel for finite signals was chosen to be either the sinc or the Gaussian. An annihilating filter approach led to an elegant algebraic solution via polynomial root finding and least squares. The authors alluded to the noisy case and suggested the use of the Singular Value Decomposition (SVD) for dealing with noisy samples. We will show that, in fact, this method is ill-conditioned because root-finding is itself not at all robust to noise. Thus it is not amenable to practical implementations, for instance on an ADC\@.

Subsequently, Dragotti {\it et al.}~\cite{Dra07} examined acquisition of the same signals with an eye toward implementability of the sampling kernel. Instead of using the sinc and Gaussian kernels (which do not have compact support), the authors limited the choice of kernels to functions satisfying the Strang--Fix conditions~\cite{StrangFix} ({\it e.g.} splines and scaling functions), exponential splines~\cite{Uns05} and functions with rational Fourier transforms. They combined the moment-sampling and annihilating filter approaches to solve for the parameters. In our work, however, we will continue to use the Gaussian as our sampling kernel. We believe that, even though the Gaussian has infinite support, it can be well approximated by its truncated version. Hence, we can still draw insights from the analysis of using Gaussian filters and the subsequent reconstruction of the signal from its noisy samples $y[n]$. More importantly, unlike with previous approaches, the sampling kernel plays no fundamental role in the reconstruction algorithm.  We use the Gaussian kernel because of its prominence in earlier work and the intuitiveness of its information spreading properties.

Maravic and Vetterli~\cite{Mar05} and Ridolfi {\it et al.}~\cite{Rid02} proposed and solved a related problem. Instead of modeling the noise at the \emph{output}, they considered the scenario where $x(t)$, the signal in question, is corrupted by additive white noise $e(t)$.
Clearly, ${x}_e(t) = x(t) + e(t)$ does not belong to the class of signals with FRI\@. However, in~\cite{Mar05}, novel algebraic/subspace-based approaches solve the sampling problem in the Laplace domain and these methods achieve some form of optimality. In~\cite{Rid02}, various algorithms including subspace-based approaches~\cite{Sto97} (\texttt{ESPRIT} and \texttt{MUSIC}) as well as multidimensional search methods were used and comparisons were made. The authors concluded that, in the noisy signal case, the parameters can be recovered at a rate below that prescribed by the Shannon-Nyquist Theorem but at a factor above the critical rate.

\subsection{Our Contributions}
In our paper, we solve a \emph{different} problem. We model the noise as additive noise to the acquired samples $y[n]$, \emph{not} the signal $x(t)$. Besides, we use the \emph{Gaussian sampling kernel} and show that the ill-conditioning of the problem can be effectively circumvented. We demonstrate that under these conditions, we are able to estimate the parameters via a fully Bayesian approach based on Gibbs sampling (GS)~\cite{Gem84, God98}. The prior methods are essentially \emph{algebraic} while our algorithm is stochastic. As such, the maximization of the log-likelihood function, which we will derive in Section~\ref{sec:mcmc}, is robust to initialization. 

More importantly, our algorithm is not constrained to work on the Gaussian kernel. \emph{Any} kernel can be employed because the formulation of the Gibbs sampler does not depend on the specific form of the kernel $h(t)$. Finally, all the papers mentioned failed to estimate the standard deviation of the noise process $\sigma_e$. We address this issue in this paper. 

\subsection{Paper Organization}
The rest of this paper is organized as follows: In Section~\ref{sec:probdef}, we will formally state the problem and define the notation to be used in the rest of the paper. We proceed to delineate our algorithm: a stochastic optimization procedure based on Gibbs sampling, in Section~\ref{sec:mcmc}. 
We report the results of extensive numerical experiments in Section~\ref{sec:results}. In Section~\ref{sec:results}, we will also highlight some of the main deficiencies in~\cite{Vet02}, which motivate the need for new algorithms for recovering the parameters of a signal with FRI given noisy samples $y[n]$. We conclude our discussion in Section~\ref{sec:concl} and provide directions for further research.

\section{Problem Definition and Notation}
\label{sec:probdef}

The basic setup is shown in Fig.~\ref{fig:block}. As mentioned in the introduction, we consider a class of signals characterized by a finite number of parameters. In this paper, similar to~\cite{Vet02, Mar05, Dra07}, the class is the weighted sum of $K$ Diracs%
\footnote{The use of a Dirac delta simplifies the discussion. 
It can be replaced by a known pulse $g(t)$
and then absorbed into the sampling kernel $h(t)$,
yielding an effective sampling kernel $g(t) * h(t)$.}
\begin{equation}
x(t)=\sum_{k=1}^{K} c_k \delta(t-t_k). \label{eqn:fri}
\end{equation}
The signal to be estimated $x(t)$ is filtered using a Gaussian low-pass filter 
\begin{equation}
h(t)=\exp\left(-\frac{t^2}{2\sigma_h^2}\right) \label{eqn:kernel}
\end{equation}
with width $\sigma_h$ to give the signal $z(t)$. Even though $h(t)$ does not have compact support, it can be well approximated by a truncated Gaussian, which does have compact support. The filtered signal $z(t)$ is sampled at rate of $1/T$ seconds to obtain $z[n]=z(nT)$ for $n=0,\, 1,\, \ldots,\, N-1$. Finally, additive white Gaussian noise (AWGN) $e[n]$ is added to $z[n]$ to give $y[n]$.
Therefore, the whole acquisition process from $x(t)$ to $\{y[n]\}_{n=0}^{N-1}$ can be represented by the model $\calM$
\begin{equation}
\calM: \quad y[n]=\sum_{k=1}^{K} c_k \exp\left(-\frac{(nT-t_k)^2}{2\sigma_h^2}\right)  +e[n]
\label{eqn:model}
\end{equation}
for $n=0,\, 1,\, \ldots,\, N-1$. The amount of noise added is a function of $\sigma_e$. We define the signal-to-noise ratio (SNR) in dB as
$$
\SNR \mathdef 10\log_{10} \left( \frac{\sum_{n=0}^{N-1} |z[n]|^2}{\sum_{n=0}^{N-1} |z[n]-y[n]|^2}\right) \,\, \mbox{dB}.
$$
In the sequel, we will use boldface to denote vectors. In particular, 
\begin{eqnarray}
\by&=&[y[0],\, y[1],\, \dots ,\, y[N-1]]\transpose , \\
\bc&=&[c_1 ,\, c_2 ,\, \dots ,\, c_K]\transpose ,\\
\bt&=&[t_1 ,\, t_2 ,\, \dots ,\, t_K]\transpose.
\end{eqnarray}
We will sometimes use $\btheta=\{ \bc, \bt, \sigma_e\}$ to denote the complete set of decision variables. We will be measuring the performance of our reconstruction algorithms by using the normalized reconstruction error 
\begin{equation}
\calE \mathdef \frac{\int_{-\infty}^{\infty} |z_{est}(t) - {z}(t)|^2\, dt}{\int_{-\infty}^{\infty} |z(t) |^2 \,dt},
\label{eqn:reconerr}
\end{equation}
where $z_{est}(t)$ is the reconstructed version of $z(t)$. By construction $\calE\ge 0$ and the closer $\calE$ is to 0, the better the reconstruction algorithm. In sum, the problem can be summarized as: {\it Given $\by=\{y[n]\, | \, n=0,\, \dots\,, N-1\}$ and the model $\calM$, estimate the parameters $\{c_k, t_k\}_{k=1}^K$ to minimize $\calE$. Also estimate the noise variance $\sigma_e^2$.} 
\section{Presentation of the Gibbs sampler}
\label{sec:mcmc}
In this section, we will describe the stochastic optimization procedure based on Gibbs sampling to estimate $\btheta=\{\bc, \bt, \sigma_e\}$. 
\subsection{Gibbs Sampling (GS)}

Markov chain Monte Carlo (MCMC) in the form of the Gibbs sampler, and the Metropolis-Hastings algorithm allows any distribution to be simulated on a finite-dimensional state space specified by any conditional density. The Gibbs sampler was first studied by the statistical physics community \cite{Met53} and then later in the statistics community \cite{Gem84, Has70, Fitz99}. The basis for Gibbs sampling is the Hammersley-Clifford theorem \cite{Ham70} which states that given the data $\by$, the conditional densities $p_i(\theta_i | \btheta_{\{j\neq i\}}, {\bf y}, \calM)$ contain sufficient information to produce samples from the joint density $p(\btheta | {\bf y}, \calM)$. Furthermore, the joint density can be directly derived from the conditional densities. 

Gibbs sampling has been used extensively and successfully in image \cite{Gem84} and audio restoration \cite{God98}. The Gibbs sampler is presented here to estimate $\btheta=\{\bc, \bt, \sigma_e\}$. To simulate our Gibbs sampler, we use the i.i.d.\ Gaussian noise assumption and the model in (\ref{eqn:model}) to express the log-likelihood of the parameters given the observations as:
\begin{eqnarray}
\lefteqn{\log p(\bc, \bt, \sigma_e\, |\, \by, \calM)} \nonumber \\
 & \propto & -(N+1)\log (\sigma_e )  \nonumber \\
 & & - {\textstyle \frac{1}{2\sigma_e^2}} \sum_{n=0}^{N-1} \left[ y[n] -\sum_{k=1}^K c_k \exp\left(-{\textstyle \frac{(nT-t_k)^2}{2\sigma_h^2}} \right) \right]^2.
\label{eqn:likeli}
\end{eqnarray}
A Jeffrey's (improper) non-informative prior has been assigned to the standard deviation of the noise such that 
\begin{equation}
p(\sigma_e) \propto \frac{1}{\sigma_e}.
\label{eqn:jeff}
\end{equation}
In the Gibbs sampling algorithm, as soon as a variate is drawn, it is inserted immediately into the conditional p.d.f.\ and it remains there until being substituted in the next iteration. This is shown in the following algorithm.\footnote{For brevity, the dependence on the model $\calM$ is omitted from the conditional density expressions.} \\
\begin{algorithmic}[H]
\REQUIRE{$\by, I, I_{b},\bm{\theta}^{(0)} = \{ \bc^{(0)}, \bt^{(0)}, \sigma_e^{(0)} \}$ }
\FOR{$i \leftarrow$ 1 : $I+I_{b}$}
\STATE $c_1^{(i)}\sim p(c_1|c_2^{(i-1)}, c_3^{(i-1)}, \dots,  c_K^{(i-1)}, \bt^{(i-1)}  \sigma_e^{(i-1)}, \by)$
\STATE $c_2^{(i)}\sim p(c_2|c_1^{(i)},c_3^{(i-1)}, \dots,  c_K^{(i-1)}, \bt^{(i-1)}  \sigma_e^{(i-1)}, \by)$
\STATE $\vdots\quad\sim\quad \vdots\quad$
\STATE $c_K^{(i)} \sim p(c_K|c_1^{(i )}, c_2^{(i)}, \dots,  c_{K-1}^{(i)},\bt^{(i-1)},   \sigma_e^{(i-1)}, \by) $
\STATE $t_1^{(i)}\sim p(t_1| \bc^{(i)},  t_2^{(i-1)},  t_3^{(i-1)}, \dots,  t_K^{(i-1)}, \sigma_e^{(i-1)}, \by) $
\STATE $t_2^{(i)}\sim p(t_2| \bc^{(i)},  t_1^{(i)}, t_3^{(i-1)}, \dots,  t_K^{(i-1)}, \sigma_e^{(i-1)}, \by) $
\STATE $\vdots\quad\sim\quad \vdots\quad$
\STATE $t_K^{(i)}\sim p(t_K|\bc^{(i)}, t_1^{(i)},  t_2^{(i)},\dots,  t_{K-1}^{(i)}, \sigma_e^{(i-1)}, \by)$
\STATE $\sigma_e^{(i)} \sim p(\sigma_e| \bc^{(i)}, \bt^{(i)}, \by)$ 
\ENDFOR
\STATE Compute $\hat{\btheta}_{\mathrm{MMSE}}$ using \eqref{eqn:MMSE}
\STATE {\bf return} $\hat{\btheta}_{\mathrm{MMSE}}$
\end{algorithmic}
In the algorithm, $\vartheta \sim \bar{p}(\cdot)$ means that $\vartheta$ is a random draw from $\bar{p}(\cdot)$. The superscript number $(i)$ denotes the current iteration. After $I_b$ iterations\footnote{$I_b$ is also commonly known as the \emph{burn-in} period in the Gibbs sampling and MCMC literature~\cite{God98}.} the Markov chain approximately reaches its stationary distribution $p(\btheta|\by, \calM)$. Minimum mean squared error (MMSE) estimates can then be approximated by taking averages of the samples from the next $I$ iterations $\{\btheta^{(I_{b}+1)},\, \btheta^{(I_{b}+2)},\, \dots,\, \btheta^{(I_{b}+I)} \}$, {\it i.e.},
\begin{equation}
\hat{\btheta}_{\mathrm{MMSE}}= \int \btheta\, p(\btheta|\by, \calM) \, d\btheta \approx \frac{1}{I} \sum_{i=I_{b}+1}^{I_{b}+I} \btheta^{(i)}. 
\label{eqn:MMSE}
\end{equation}

\subsection{Presentation of the Posterior Densities in the GS}
We will now derive the conditional densities.
In the sequel, we will use the notation $\btheta_{-\ell}$ to denote the set of parameters excluding the $\ell$th parameter. 
It follows from Bayes' theorem that 
\begin{equation}
p(\theta_\ell |\btheta_{-\ell}, \by, \calM)\propto p(\by  |\btheta,  \calM) \, p(\btheta).
\end{equation}
Thus, the required conditional distributions are proportional to the \emph{likelihood} of the data times the \emph{priors} on the parameters. The likelihood function of $\by$ given the model is given in \eqref{eqn:likeli} from the Gaussian noise assumption. Thus, we can calculate the posterior distributions of the parameters given the rest of the parameters. The parameters conditioned on are taken as constant and can be left out of the posterior. We will sample from these posterior densities in the GS iterations as shown in the above algorithm.

We will now proceed to present the posterior densities. The derivations are provided in the Appendix.
\subsubsection{Sampling $c_k$} 
$c_k$ is sampled from a Gaussian distribution given by
\begin{equation}
p(c_k|\btheta_{-c_k}, \by, \calM) = \calN \left(c_k; -\frac{\beta_k}{2\alpha_k}, \frac{1}{2\alpha_k}\right),
\end{equation}
where
\begin{eqnarray}
\alpha_k&\mathdef&\frac{1}{2\sigma_e^2} \sum_{n=0}^{N-1} \exp\left(-\frac{(nT-t_k)^2}{\sigma_h^2}\right), \label{eqn:alpha}\\
\beta_k& \mathdef &\frac{1}{\sigma_e^2}   \sum_{n=0}^{N-1}  \exp\left(-\frac{(nT-t_k)^2}{2\sigma_h^2}\right) \nonumber \\
& & \cdot
\left\{ \sum_{\substack{k'=1\\ k'\ne k}}^{K } c_{k'}  \exp\left(-\frac{(nT-t_{k'})^2}{2\sigma_h^2}\right)  -y[n]\right\}.  \label{eqn:beta}
\end{eqnarray}
It is easy to sample from Gaussian densities when the parameters $(\alpha_k, \beta_k)$ have been determined.
\subsubsection{Sampling $t_k$} 
$t_k$ is sampled from a distribution of the form
\begin{eqnarray}
\lefteqn{p(t_k|\btheta_{-t_k}, \by, \calM)} \nonumber \\
& \propto & \exp\Bigg[ -\frac{1}{2\sigma_e^2} \sum_{n=0}^{N-1} \gamma_k  \exp\left(-\frac{(nT-t_k)^2}{\sigma_h^2} \right) \nonumber \\
& & \qquad + \; \nu_k\exp\left(-\frac{(nT-t_k)^2}{2\sigma_h^2}\right) \Bigg]
\label{eqn:tk}
\end{eqnarray}
where
\begin{eqnarray}
\gamma_k & \mathdef & c_k^2, \label{eqn:gamma} \\
 \nu_k & \mathdef & 2c_k \left\{ \sum_{\substack{k'=1\\ k'\ne k}}^{K} c_{k'}  \exp\left(-\frac{(nT-t_{k'})^2}{2\sigma_h^2}\right)-y[n] \right\} . \label{eqn:nu}
\end{eqnarray}
It is not straightforward to sample from this distribution. We can sample $t_k$ from a uniform grid of discrete values with probability masses proportional to~\eqref{eqn:tk}. But in practice, and for greater accuracy, we used rejection sampling~\cite{Tie94, Rob04} to generate samples $t_k^{(i)}$ from $p(t_k|\btheta_{-t_k}, \by, \calM)$. The proposal distribution $\tilde{q}(t_k)$ was chosen to be an appropriately scaled Gaussian, since it is easy to sample from Gaussians. This is shown in the following algorithm.\\
\begin{algorithmic}
\REQUIRE{$\tilde{p}(t_k) \mathdef p(t_k| \btheta_{-t_k}, \by, \calM)$}
\STATE Select $\tilde{q}(t_k)\sim\calN$ and $c$ s.t. $\tilde{p}(t_k)<c\tilde{q}(t_k)$
\STATE $u\sim\mathcal{U}(0,1)$
\REPEAT
\STATE $t_k \sim \tilde{q}(t_k)$
\UNTIL $u < \Frac{\tilde{p}(t_k)}{(c\tilde{q}(t_k))}$
\end{algorithmic} 

\subsubsection{Sampling $\sigma_e$} 
$\sigma_e$ is sampled from the `Square-root Inverted-Gamma'~\cite{Ber01} distribution $\mathcal{IG}^{-1/2} (\sigma_e; \varphi, \lambda)$\footnote{$X$ follows a `Square-root Inverted-Gamma' distribution if $X^{-2}$ follows a Gamma distribution.},
\begin{equation}
p(\sigma_e|\btheta_{-\sigma_e}, \by, \calM)
= \frac{2\lambda^{\varphi} \sigma_e^{-(2\varphi+1)}}{\Gamma(\varphi)} \exp\left(-\frac{\lambda}{\sigma_e^2}\right) \mathbb{I}_{[0,+\infty)} (\sigma_e) , 
\end{equation}
where 
\begin{eqnarray}
\varphi&\mathdef&\frac{N}{2}, \label{eqn:varphi} \\ 
\lambda& \mathdef & \frac{1}{2} \left[ y[n] -\sum_{k=1}^K c_k  \exp\left(-\frac{(nT-t_k)^2}{2\sigma_h^2} \right) \right]^{2 }. \label{eqn:lambda}
\end{eqnarray} 
Thus the distribution of the variance of the noise $\sigma_e^2$ is Inverted Gamma, which corresponds to the conjugate prior of $\sigma_e^2$ in the expression of $\calN(e; 0,\sigma_e^2)$~\cite{Ber01} and thus it is easy to sample from. In our simulations, we sampled from this density using the Matlab function \texttt{gamrnd} and applied the `Inverted Square-root' transformation

\subsection{Further Improvements via Linear Least Squares Estimation} \label{sec:llse}

\begin{figure}
  \centering
  \subfigure[Histogram of the samples of $c_1^{(i)}$]
  {
      \includegraphics[height=1.9in]{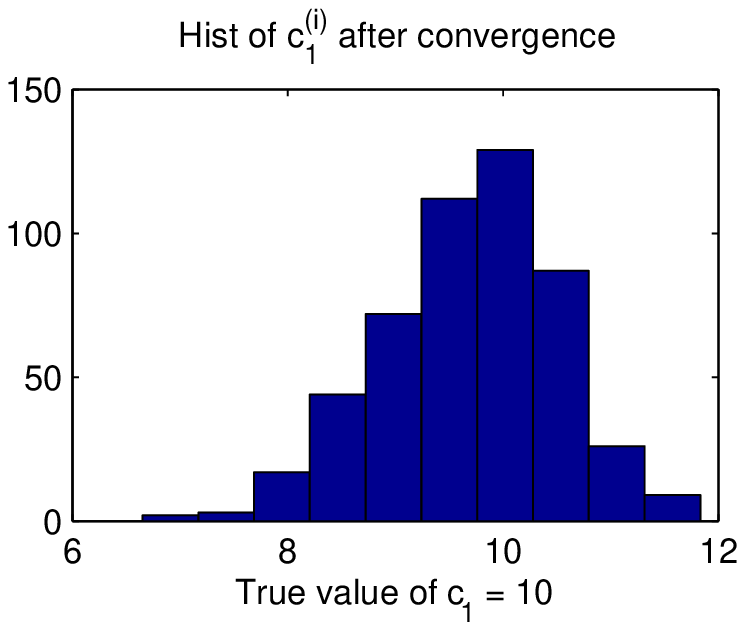}
      \label{fig:hist1}
  }      \hspace{.3in}
  \subfigure[Histogram of the samples of $t_1^{(i)}$]
  {
      \includegraphics[height=1.9in]{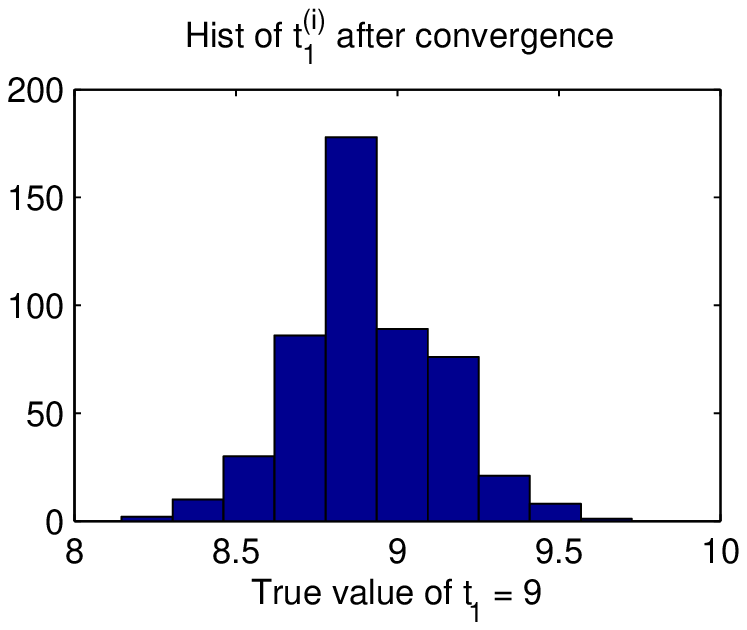}
      \label{fig:hist2}
  }  
   \caption{Note that the variance of the stationary distribution of the $t_k$s is smaller than that of the $c_k$s after convergence of the Markov chain. }
  \label{fig:hist}
\end{figure}

We can perform an additional post-processing step to improve on the estimates of $c_k$. We noted from our preliminary experiments (see Fig.~\ref{fig:hist}) that the variance of the stationary distribution of the $t_k$s is smaller than that of the $c_k$s. This results in better estimates for the locations $t_k$s as compared to the magnitudes $c_k$s. 
Now, we observe that $y[n]$, the observations, are linear in the $c_k$s once the $t_k$s are known. A natural extension to our GS algorithm is to augment our $c_k$ estimates with a linear least squares estimation (LLSE) procedure using $\by$ and the MMSE estimates of $t_k$. Eqn.\ \eqref{eqn:model} can be written as
\begin{equation}
y[n]=\sum_{k=1}^Kc_k h(nT-t_k)+e[n],\quad 0\le n\le N-1
\label{eqn:llse}
\end{equation}
with $h(t)$, the Gaussian sampling kernel, given in \eqref{eqn:kernel}. Given the set of estimates of the time locations $\{\hat{t}_k\}_{k=1}^K$, we can rewrite \eqref{eqn:llse} as a matrix equation, giving
$$
\by=\mathbf{H}\bc+\mathbf{e},
$$
where $[\mathbf{H}]_{nk}=h(nT-\hat{t}_k)$ and $1\le n\le N$, $1\le k\le K$. We now minimize the square of the residual $\|\mathbf{e} \|^2 =\|\mathbf{H}\bc-\by \|^2$, giving the normal equations $\mathbf{H}\transpose\mathbf{H}\bc=\mathbf{H}\transpose\by$ and the least squares solution~\cite{Strang}
\begin{equation}
\hat{\bc}_{\mathrm{LS}}=(\mathbf{H}\transpose\mathbf{H})^{-1}\mathbf{H}\transpose\by.
\label{eqn:LLSE}
\end{equation}
From our experiments, we found that, in general, using $\hat{\bc}_{\mathrm{LS}}$ as estimates for the magnitudes of the impulses provided a lower reconstruction error $\calE$.

\section{Numerical Results and Experiments}
\label{sec:results}
In this section, we will first review the annihilating filter and root-finding method algorithm for solving for the parameters of a signal with FRI\@. This algorithm provides a baseline for comparison. Then we will provide extensive simulation results to validate the accuracy of the algorithm we proposed in Section~\ref{sec:mcmc}.

\subsection{Problems with Annihilating Filter and Root-Finding}

\begin{table}
\centering
\begin{tabular}{|c||c|c|c|c|}
\hline
Param. & $K$  & $\sigma_e$& $N$&$\SNR$ 
\\ \hline
Value & 5& 0 and $10^{-6}$&30  & $\infty$ and 137 dB  \\ \hline
\end{tabular}
\caption{Parameter values for demonstration of the annihilating filter and root-finding algorithm. }
\label{tab:sd}
\end{table}
\begin{figure}
  \centering
  \subfigure[The annihilating filter approach reconstructs the signal exactly in the noiseless scenario.]
  {
      \includegraphics[height=1.9in]{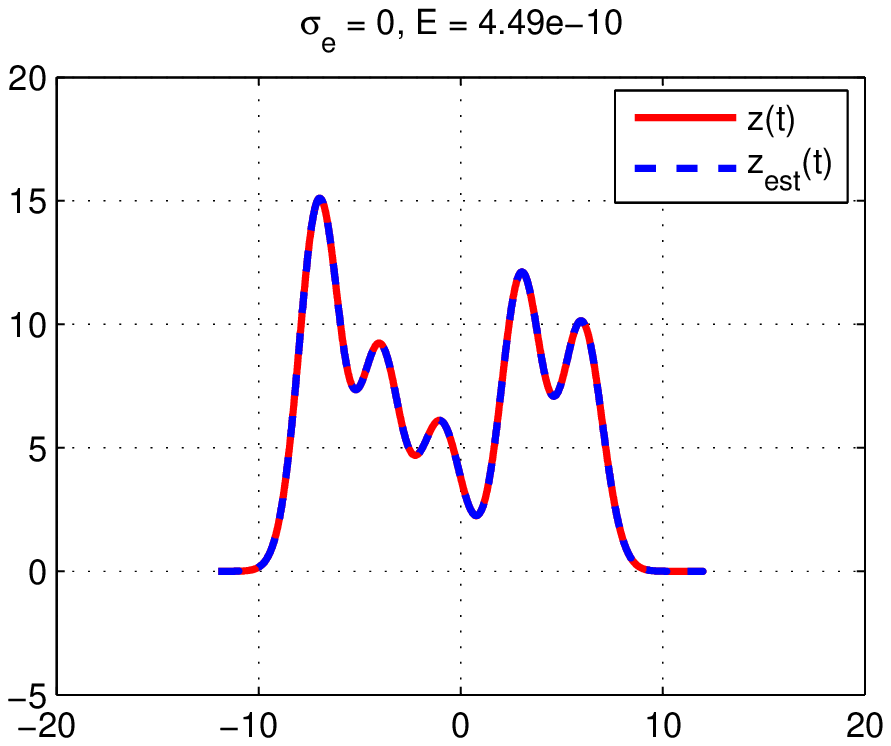}
      \label{fig:ann1}
  }      \hspace{.3in}
  \subfigure[The reconstruction completely breaks down when noise of a small standard deviation $\sigma_e=10^{-6}$ (SNR=137 dB) is added.]
  {
      \includegraphics[height=1.9in]{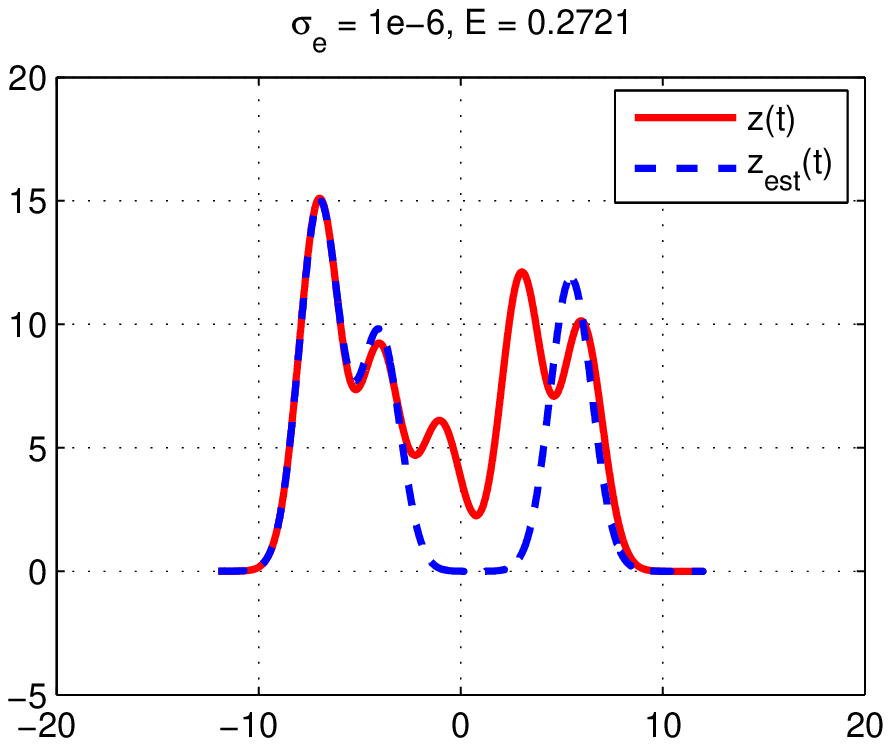}
      \label{fig:ann2}
  }  
   \caption{Demonstration of the annihilating filter/root-finding approach. }
  \label{fig:ill}
\end{figure}

In~\cite{Vet02}, Vetterli {\it et al.} introduced the concept of a class of signals with a finite rate of innovation. For signals of the form \eqref{eqn:fri} and certain sampling kernels, the annihilating filter was used as a means to locate the $t_k$ values. Subsequently a least squares approach yielded the weights $c_k$. It was shown that in the noiseless scenario, this method recovers the parameters exactly (see Fig.~\ref{fig:ill}(a)). For completeness, we will briefly outline their method here. Denoting the noiseless samples by $z[n]$, \eqref{eqn:model} can be written as
\begin{equation}
p[n]=\sum_{k=1}^K a_k u_k^n, \quad n=0,\,1,\, \dots,\, N-1,
\end{equation}
with the identifications 
\begin{eqnarray}
p[n]&=&\exp(n^2 T^2/(2\sigma_h^2))z[n], \label{eqn:expweight}\\
a_k &=&c_k\exp(-t_k^2/(2\sigma_h^2)),\label{eqn:af}\\
u_k&=&\exp(t_k T/\sigma_h^2).
\end{eqnarray}
Now, since $p[n]$ is a linear combination of exponentials, we find the annihilating filter $a[n]$ such that 
$$
a[n]\ast p[n]=\sum_{\ell=0}^K a[\ell]p[n-\ell] =0, \quad \forall n\in\mathbb{Z}.
$$
This can be written in matrix/vector form as $\bP \ba=\mathbf{0}$. This system will admit a solution when rank$(\bP)=K$. In practice this is solved using an SVD where $\bP=\bU\bm{\Sigma} \bV^T$ and $\ba=\bV \be_{K+1}$ and $\be_{K+1}$ is a length-($K+1$) vector with 1 in position $(K+1)$ and 0 elsewhere. Now, once the coefficients $a[n]$ are found, the values $u_k$ are simply the roots of the filter 
$$
A(z)=\sum_{n=0}^K a[n]z^{-n}.
$$
The $t_k$s can then be determined from \eqref{eqn:af} and the solution for the $c_k$s essentially parallels the development in Section~\ref{sec:llse}.

In the same paper, it was suggested that to deal with the noisy samples, we can minimize $\|\bP\ba\|$, in which case, $\ba$ is the eigenvector that corresponds to the smallest eigenvalue of $\bP\transpose \bP$. Here, we argue that this method is inherently ill-conditioned and thus not robust to noise.

\begin{enumerate}
\item Firstly, minimizing $\|\bP\ba\|$ involves finding the eigenvector $\mathbf{v}_{1}$ that corresponds to the largest eigenvalue $\lambda_{1}$. Because computing eigenvalues and eigenvectors are essentially root-finding operations, this is ill-conditioned. 
\item Secondly, even if the vector $\ba= \mathbf{v}_{1}$ can be found, the zeros of the filter $A(z)$ have to be found. This again involves root finding, which is ill-conditioned. 
\item Finally, from \eqref{eqn:expweight}, any noise added to $z[n]$ will be exponentially weighted in the observations $p[n]$. We feel that this is the greatest source of ill-conditioning. 
\end{enumerate}
Because of the three reasons highlighted above, there is a need to explore new algorithms for finding the parameters. In Fig.~\ref{fig:ill}, we show a simulation with the parameters as tabulated in Table~\ref{tab:sd}, but we varied the noise ($\sigma_e=10^{-6}$ gives $\SNR=137$ dB, a very low noise level). We observe from Fig.~\ref{fig:ann2} that (without oversampling) the annihilating filter and root-finding method is not robust even when a miniscule amount of noise is added.
\begin{remark}
The root-finding method is so unstable that, at times, even for low levels of noise, we obtain complex roots for the locations $\{t_k\}_{k=0}^{K-1}$. To solve this problem, we orthogonally projected the polynomial described by the filter coefficients $a[n]$ to the closest polynomial that belongs to the space of polynomials with real roots only. 
\end{remark}

\subsection{Performance of our GS Algorithm}
Clearly, the annihilating filter/root-finding algorithm is not robust to noise. We have suggested an alternative reconstruction algorithm in Section~\ref{sec:mcmc}, and in this section, we will present our results on several synthetic examples.\footnote{All the code, written in MATLAB, can be found at the first author's homepage \url{http://web.mit.edu/vtan/frimcmc}.}

\subsubsection{Initial Demonstration}
\begin{figure}
  \centering
  \subfigure[Evolution of the $c_k$s]
  {
      \includegraphics[height=1.9in]{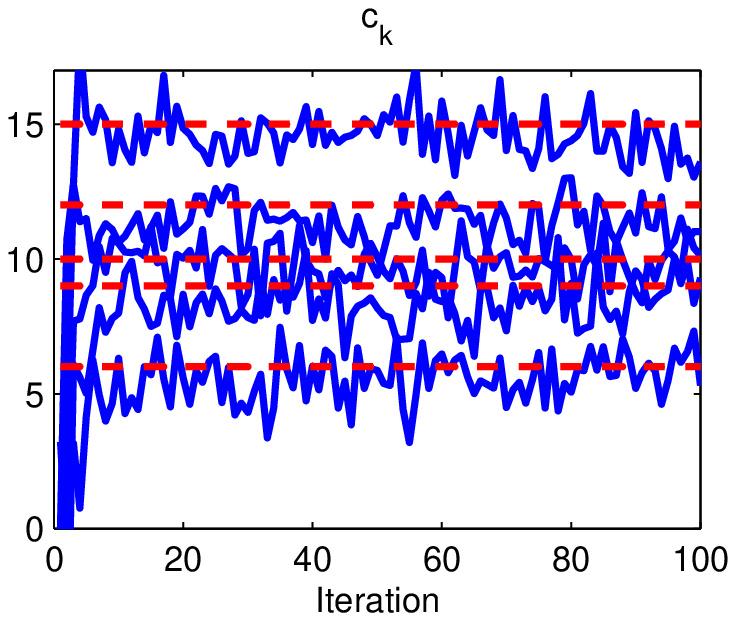}
      \label{fig:ck}
  }      \hspace{.3in}
  \subfigure[Evolution of the $t_k$s]
  {
      \includegraphics[height=1.9in]{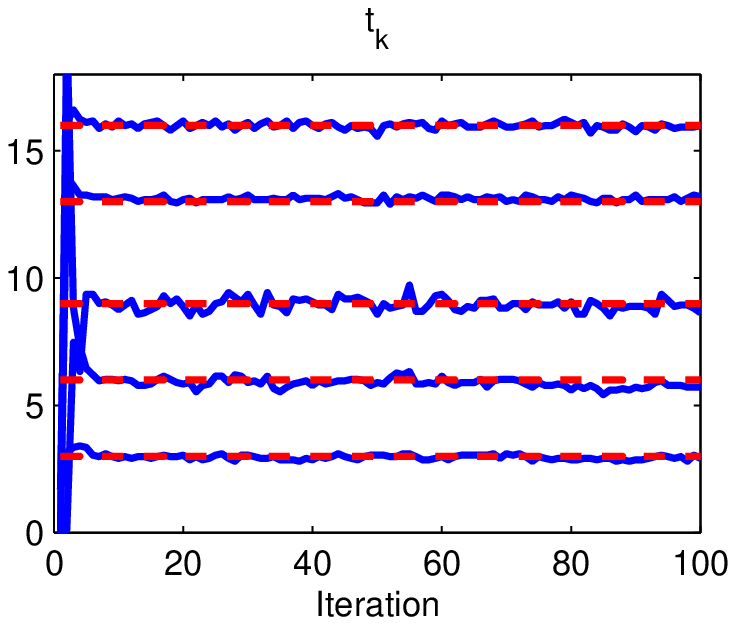}
      \label{fig:tk}
  }\\
  \subfigure[Evolution of the $\sigma_e$]
  {
      \includegraphics[height=1.9in]{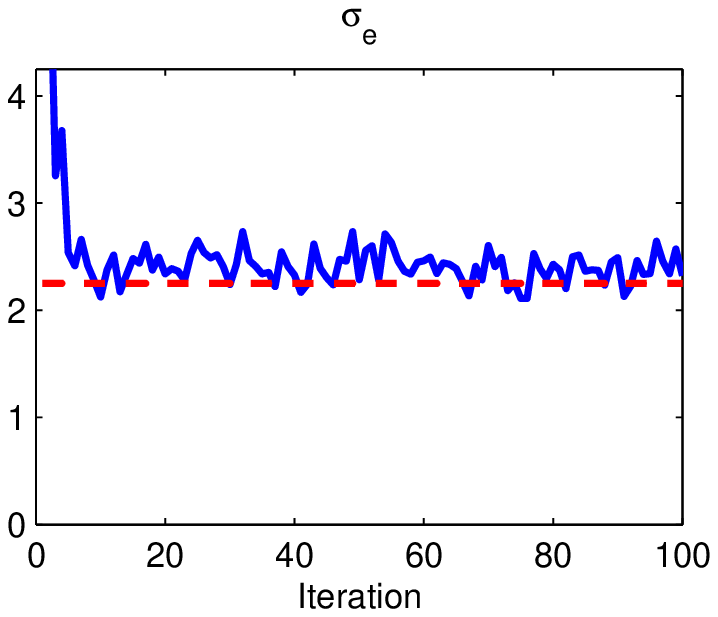}
      \label{fig:sigmae}
  }      \hspace{.3in}
  \subfigure[Reduction of the (negative) log-likelihood $-\log p(\bc, \bt, \sigma_e\, |\, \by, \calM) $]
  {
      \includegraphics[height=1.9in]{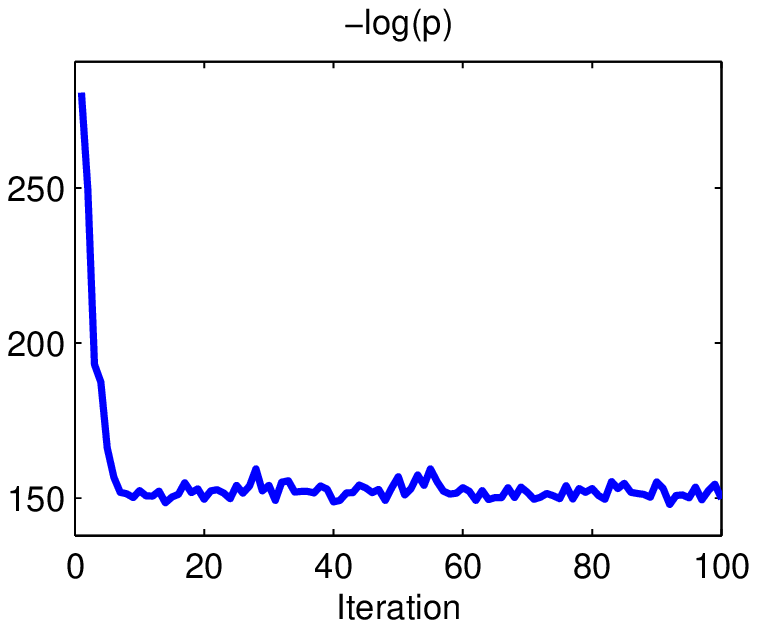}
      \label{fig:logp}
  }
  \caption{Evolution of the GS algorithm. The iterates of the parameters $\{c_k, t_k\}_{k=1}^K$ and $\sigma_e$ are shown. The true values are indicated by the broken red lines. In Fig.~\ref{fig:logp}, we see that the negative log-likelihood converges to the global minimum in fewer than 20 iterations for this problem size ($K=5$).}
  \label{fig:gsalgo}
\end{figure}

\begin{figure}
\centering
\includegraphics[height=1.9in]{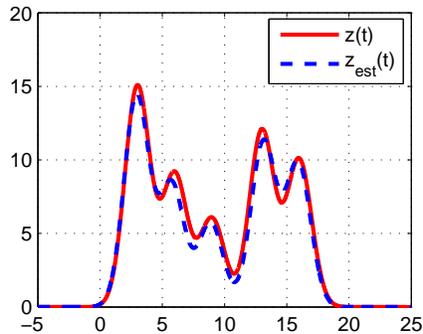}
\caption{Comparison between $z(t)$ and $z_{est}(t)$ using the GS algorithm. For this run, $\calE=0.0072$.}
\label{fig:recon}
\end{figure}

To demonstrate the evolution the Gibbs sampler, we performed an initial experiment and chose the parameters to be those in Table~\ref{tab:sd}, with the exception that the noise standard deviation was increased to $\sigma_e=2.5$, giving an SNR of $10.2$ dB\@. We plot the iterates in Fig.~\ref{fig:gsalgo}. The true filtered signal $z(t)$ and its estimate $z_{est}(t)$ are plotted in Fig.~\ref{fig:recon}. We note the close similarity between $z(t)$ and $z_{est}(t)$.

We observe that the sampler converges in fewer than 20 iterations for this run, even though the parameter values were initialized far from their optimal values. We emphasize that as GS is essentially a stochastic optimization procedure (not unlike Simulated Annealing or Genetic Algorithms), it is insensitive to the choice of starting point $\btheta^{(0)}$.  The Markov Chain is guaranteed to converge to the stationary distribution after the burn-in period~\cite{Tie94}. 

\subsubsection{Further Experiments}
\begin{table}
\centering
\begin{tabular}{|c||c|c|c|c|c|c|c|}
\hline
Param. & $K$ &  $\sigma_e$& $N$&$\SNR$ & $\calE$
\\ \hline 
Expt A & 5 &  1.5:0.25:3.0 & 50:25:150  & Fig~\ref{fig:snr1} & Fig~\ref{fig:gibbs1}   \\ \hline
Expt B & 10 &  3.0:0.50:6.0 & 100:50:250  & Fig~\ref{fig:snr2}& Fig~\ref{fig:gibbs2}   \\ \hline
\end{tabular}
\caption{Parameter values for numerical simulations. }
\label{tab:numsim}
\end{table}
\begin{figure}
  \centering
  \subfigure[SNR (dB) against $\sigma_e$ for Expt A ($K=5$).]
  {
      \includegraphics[height=1.9in]{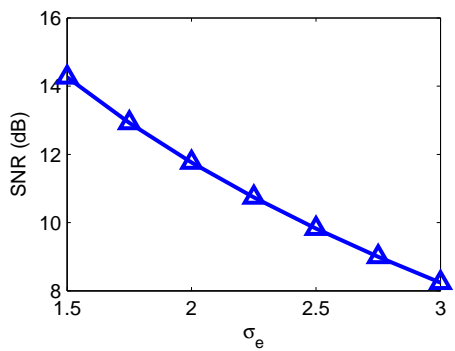}
      \label{fig:snr1}
  }      \hspace{.3in}
  \subfigure[SNR (dB) against $\sigma_e$ for Expt B ($K=10$).]
  {
      \includegraphics[height=1.9in]{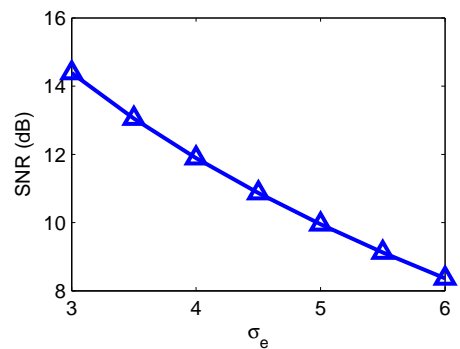}
      \label{fig:snr2}
  }  \caption{SNR (dB) against $\sigma_e$ for the two experiments. }
  \label{fig:snrplots}
\end{figure}
\begin{figure}
  \centering
  \subfigure[Errors $\calE$ against $\sigma_e$ for Expt A ($K=5$).]
  {
      \includegraphics[height=1.9in]{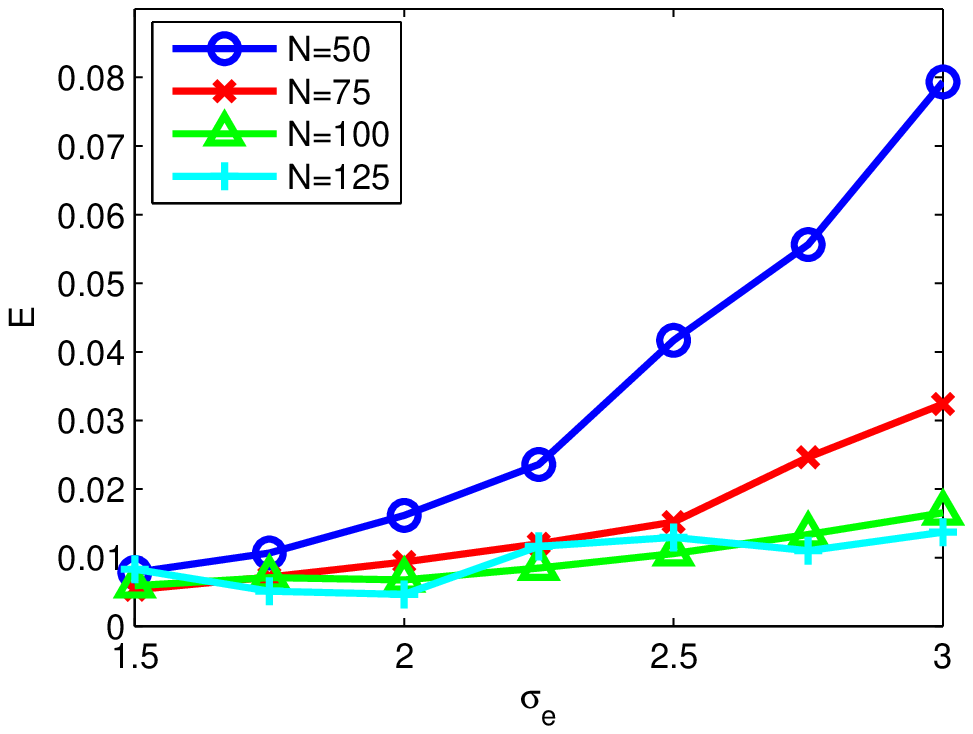}
      \label{fig:gibbs1}
  }      \hspace{.3in}
  \subfigure[Errors $\calE$ against $\sigma_e$ for Expt B ($K=10$).]
  {
      \includegraphics[height=1.9in]{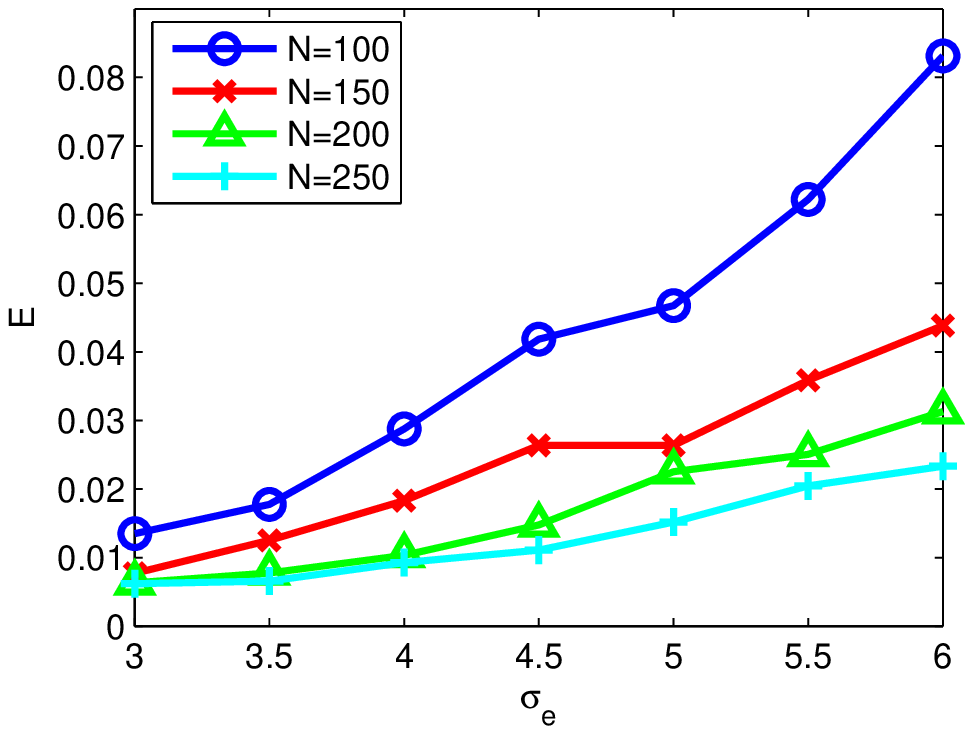}
      \label{fig:gibbs2}
  }
  \caption{Plots of $\calE$ against $\sigma_e$ for various oversampling factors and problem sizes.}
  \label{fig:gs}
\end{figure}
To further validate our algorithm, we performed extensive simulations (Expts A and B) on two different problem sizes to validate our algorithm. For consistency, each experiment was repeated using 100 different random seeds and the means of $\calE$ [cf.\ \eqref{eqn:reconerr}] taken. The parameters are chosen according to Table~\ref{tab:numsim}. The unknown parameters were initialized as $\bc^{(0)}=\bt^{(0)}=[0, \dots, 0]$ and $\sigma_e^{(0)}= 0.01$. The results for Expts A and B are shown in Fig.~\ref{fig:gibbs1} and~\ref{fig:gibbs2} respectively. We noted from these experiments that:

\begin{itemize}
\item The GS algorithm is insensitive to initialization. It \emph{always} finds approximately optimal estimates from any starting point because the Markov chain provably converges to the stationary distribution~\cite{Tie94}.
\item The LLSE post-processing step in the GS algorithm reduces the reconstruction error $\calE$. This is a consequence of using the (more accurate) $t_k$s from the sampler to estimate the $c_k$s via LLSE, instead of using the $c_k$s from the sampler directly. 
\item From the two plots in Fig.~\ref{fig:gs}, we observe that, if the problem size doubles (from $K=5$ to $K=10$), with corresponding doubling of $(\sigma_e,N)$, $\calE$ remains approximately constant. This insures scalability of the algorithm. For example, $\calE(K=5, \sigma_e=2.5, N=50)\approx \calE(K=10, \sigma_e=5.0, N=100)\approx 0.045$. 
\item The noise standard deviation $\sigma_e$ can be estimated accurately in the GS algorithm as shown in Fig.~\ref{fig:sigmae}. This may be important in some applications. 
\end{itemize}
To conclude, even though the annihilating filter approach~\cite{Vet02} is more computationally efficient than our algorithms, it is certainly not amenable to scenario where noisy samples are acquired. 

\section{Conclusions}
\label{sec:concl} 
In this paper, we addressed the problem of reconstructing a signal with FRI given noisy samples. We showed that it is possible to circumvent some of the problems of the annihilating filter and root-finding approach~\cite{Vet02}. We introduced the Gibbs sampling algorithm. From the performance plots, we observe that GS performs very well as compared to the annihilating filter method, which is not robust to noise. 

Perhaps the most important observation we made is the following: The success of the \emph{fully Bayesian} GS algorithm does not depend on the choice of kernel $h(t)$. The formulation of the GS does not depend on the specific form of $h(t)$. In fact, we used a Gaussian sampling kernel to illustrate that our algorithm is not restricted to the classes of kernels considered in~\cite{Dra07}. 

A natural extension to our work here is to assign structured priors to $\bc$, $\bt$ and $\sigma_e$. These priors can themselves be dependent on their own set of \emph{hyperparameters}, giving a hierarchical Bayesian formulation. In this way, there would be greater flexibility in the parameter estimation process. We can also seek to improve on the computational load of the algorithms introduced here. Another interesting research direction is to examine the feasibility of using the subspace-based approaches~\cite{Mar05} to solve the problem of acquired samples that are noisy. 

A question that remains is: \emph{How well can real-world signals (including natural images) be modeled as signals with FRI?} We believe the answer will have profound ramifications for areas such as sparse approximation~\cite{DVDD98} and compressed sensing~\cite{Don06,CT06}.

\appendix[Derivation of the Conditional Densities]

For brevity, we define
$$
g_{nk} \mathdef h(nT-t_k)= \exp\left( -\frac{(nT-t_k)^2}{2\sigma_h^2}\right).
$$
We start from the log-likelihood
of the parameters $\btheta$ given the data $\by$ and model $\calM$
[cf.\ \eqref{eqn:likeli}].
To obtain $p(c_k|\btheta_{-c_k}, \by, \calM)$, we treat the other parameters $\btheta_{-c_k}$ as constant, giving $\log p(c_k|\btheta_{-c_k}, \by, \calM)$ proportional to
$$
-\frac{1}{2\sigma_e^2} \sum_{n=0}^{N-1} \left[ c_k^2 g_{nk}^2 +2c_k g_{nk}   \left( \sum_{\substack{k'=1\\ k'\ne k}}^{K} c_{k'} g_{nk'}-y[n] \right) \right].
$$
Comparing this expression in $c_k$ to the Gaussian distribution with mean $\mu$ and variance $\sigma^2$,
$$
\log p(c_k; \mu, \sigma^2) \propto -\frac{1}{2\sigma^2} (c_k-\mu)^2,
$$
and equating coefficients, we obtain \eqref{eqn:alpha} and \eqref{eqn:beta}. The distribution $p(t_k|\btheta_{-t_k}, \by, \calM)$ can be obtained similarly and is omitted. Finally for the noise standard deviation $\sigma_e$, 
$$
\log p(\sigma_e|\btheta_{-\sigma_e}, \by, \calM) \propto -(N+1)\log(\sigma_e)-\frac{\lambda}{\sigma^2},
$$
where $\lambda$ is defined in \eqref{eqn:lambda}. Taking the antilog on both sides yields
$$
 p(\sigma_e|\btheta_{-\sigma_e}, \by, \calM)\propto \sigma_e^{-(N+1)} \exp\left( -  \frac{\lambda}{\sigma^2}\right),
$$
which is the `Square-root Inverted-Gamma' distribution with parameters given by \eqref{eqn:varphi} and \eqref{eqn:lambda}. All the densities have been derived. 

\bibliographystyle{IEEEtran}
\bibliography{projectbib}

\end{document}